# Detecting Neural Trojans Through Merkle Trees

*Joshua Strubel*

*Savannah River National Laboratory[1], Joshua.strubel@srnl.doe.gov*

*Abstract* - Deep neural networks are utilized in a growing number of industries. Much of the current literature focuses on the applications of deep neural networks without discussing the security of the network itself. One security issue facing deep neural networks is neural trojans. Through a neural trojan, a malicious actor may force the deep neural network to act in unintended ways. Several potential defenses have been proposed, but they are computationally expensive, complex, or unusable in commercial applications. We propose Merkle trees as a novel way to detect and isolate neural trojans.

## I. Introduction

A deep neural network (DNN) is a neural network organized into different layers of neurons. A common DNN architecture includes multiple layers consisting of input, hidden, and output layers respectively. The input layer is responsible for receiving the input of training, testing, or validation data. In the context of a DNN classifier, each neuron in the output layer corresponds to an individual classification label. Appended to the output layer is an activation function, often a SoftMax function, which can be modeled by (1.1).

$$\text{Equation 1.1 } \sigma(\vec{z}) \frac{e^{z_i}}{\sum_{j=1}^{K} e^{z_j}}$$

$$\text{where } \vec{z} = input\ vector,$$

$$e^{z_i} = exponential\ function\ for\ input\ vector,$$

$$K = Number\ of\ classes,$$

$$e^{z_j} = exponential\ function\ for\ output\ vector$$

The largest $e^{z_j}$ is the selected output label of the classifier. Because of the inherent architecture of DNNs, it has been shown that a malicious actor through neural trojans can force a DNN to behave in a manner of the malicious actor's choosing [1], [2]. A neural trojan, much like a traditional trojan, embeds itself within the neural network and masquerades as benign parameters within the network. A malicious actor may alter either the hardware or the software to insert a neural trojan into the DNN [1], [2]. The software neural trojan often takes the form of either a single-bias attack or a gradient-descent style attack [1]. A gradient-descent style attack is more complex and resembles the training process of an uncompromised DNN [1]. Through gradient descent, a malicious actor will force the DNN to behave erroneously on a small, selected subset of data [1]. Inserting this type of neural trojan without detection is difficult; but if successfully inserted, is very difficult to detect and remediate in the future.

The single-bias attack is much simpler and easier to insert. A single-bias attack works by adding a bias, $\varepsilon$, to a single neuron [1]. Because of the precision of this attack, it is less noisy and more difficult to detect. The effectiveness of this attack consists of the ability of the malicious actor to force the desired behavior of the DNN without overly degrading the model's performance, drawing attention to the presence of the neural trojan.

Because of the vulnerability of DNNs to neural trojans, there have been several proposed defenses. These defenses include monitoring neuron's activation on select input [3], comparing DNN's behavior to an ensemble of similar benign DNNs [4], and lastly, custom checkpoints to compare the hashes of model parameters [5]. Each defense has its merits, but none of them consists of a commercially viable defense. There is a need for a neural trojan defense that is both computationally inexpensive and easy to implement. We propose Merkle trees as a novel and computationally inexpensive and easy to implement defense against neural trojans

In this paper, we will discuss the theory behind both the single-bias neural trojan and the proposed Merkle tree defense in the methods section. We will then demonstrate the effectiveness of this defense in the section on experimentation, which will be followed by a discussion of the results. Lastly, we will conclude with a discussion of the implications and importance of the results.

---

[1] This work was supported by the Laboratory Directed Research and Development (LDRD) program within the Savannah River National Laboratory (SRNL). This document was prepared in conjunction with work accomplished under Contract No. 89303321CEM000080 with the U.S. Department of Energy (DOE) Office of Environmental Management (EM).

## II. Methods

Because of their simplicity, precision, and difficulty to detect, we selected the single-bias attack for our research. For the rest of this paper, we will limit our discussion of neural trojans to the single-bias attack on a DNN classifier. This attack is modeled by (1.2) and (1.3).

$$Equation\ 1.2\ minimize\ f_{adv}(\emptyset_n - \varepsilon)$$

$$w.r.t. |\varepsilon| \leq |L|$$

$$where\ f_{adv}(\emptyset)\ is\ the\ probability\ of\ classifying$$

$$input\ as\ the\ target\ sink\ class,$$

$$\varepsilon\ is\ a\ bias\ value, and\ L\ is\ the\ change$$

$$In\ the\ accuracy\ of\ the\ baseline\ model.$$

By adding a significantly large bias to a specific parameter, a malicious actor can force a DNN classifier to misclassify an entire sink class. The effectiveness of this attack is measured by both the probability of correctly classifying a targeted sink class and the change in the model's accuracy. A model with sudden large degradation in performance is likely to be closely examined, potentially exposing the inserted single-bias attack. A sink class as mentioned in (1.2) is defined by (1.3).

$$Equation\ 1.3\ Sink\ class\ of\ \emptyset_n = f(\emptyset_n + \varepsilon)$$

$$where\ f(\emptyset)\ is\ the\ output\ of\ a\ DNN\ classifier$$

$$with\ all\ output\ neurons\ initialized\ to\ 0,$$

$$and\ \varepsilon\ is\ a\ bias\ value\ of\ 1$$

Per (1.1), a DNN with a SoftMax function will select the output label associated with the largest neuron of the output layer. For DNN classifiers with a SoftMax function, the effect of the single-bias attack is not uniform for every layer [1]. Because of this, the attack is most effective on the output layer, which contains a neuron for each corresponding sink class. A visible representation of a single-bias attack can be seen in Figure 1.1. In this figure, we see a simplification of a single-bias attack on a DNN classifier of handwritten images. By adding a perturbation to one of the output layer parameters, the malicious actor is able to force the misclassification of the sink class of "three."

Because of the limited number of classes and neurons, detecting and locating a single-bias attack on a classifier such as an MNIST classifier is trivial; but for larger more complex deep learning models such as BERT, with its 110,000,000 parameters, detecting a single-bias attack is a considerable task. One computationally cheap and novel way of detecting the presence of a trojan would be using training checkpoints to compare the hashes of the DNN's weights [5]. This method is effective at detecting the presence of a trojan, but it does not give any indication of the location of the single-bias attack [5].

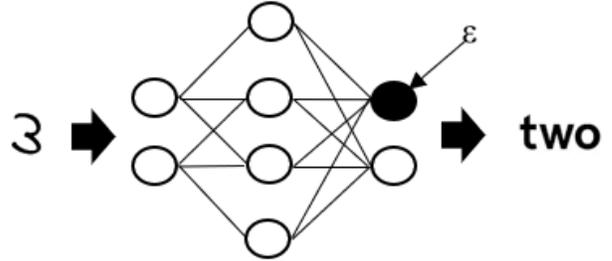

*Figure 1.1 Single-Bias Attack on DNN Classifier*

Commercial DNNs can be very large and expensive to replace. It would be commercially expedient to be able to isolate and remedy a trojaned DNN. A naïve solution would be to run through every neuron to look for modification, but this is not an efficient or scalable solution. We propose Merkle trees are an efficient solution to this problem. Merkle trees were originally made popular in the 1980s in digital signature research because the security of a Merkle-tree-based signature was not dependent upon the difficulty of a specific mathematical equation [6]. Merkle trees were revisited in the 2000s because of Bitcoin network's reliance on recorded transactions [7].

A Merkle tree is a unique data structure made up of a tree of hashes. As seen in (1.4), at the root of the tree is the root hash, which consists of a hash of its leaves and their subsequent children. Any change in any part of the data will propagate to the root hash, allowing the identification and location of where the change took place.

$$Equation\ 1.4\ Root\ Hash = MHT(\emptyset_1, \ldots, \emptyset_n) = H($$

$$H($$

$$H(H(\emptyset_1), H(\emptyset_2)), H(H(\emptyset_3), H(\emptyset_4))$$

$$),$$

$$H($$

$$H(H(\emptyset_{n-3}), H(\emptyset_{n-2})), H(H(\emptyset_{n-1}), H(\emptyset_n))$$

$$)$$

$$)$$

$$where\ \emptyset_n\ is\ an\ individual\ neuron\ within\ a\ DNN$$

In addition to their change locating capability, Merkle trees are incredibly efficient to traverse. Merkle trees have a search complexity of O(log(n)), making them ideal for searching through the neurons of large commercial DNNs. In our proposed Merkle tree of DNN neurons, as seen in Figure 1.2, the root of the tree is the root hash, and each parent in the tree is the hash of its children (neurons). This data structure will allow us both to detect a neural trojan through the root hash and then to locate the trojaned neuron efficiently by traversing the tree.

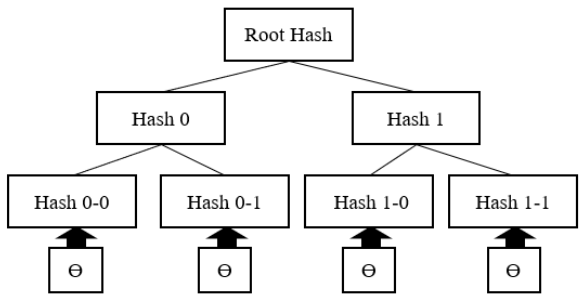

*Figure 1.2 Merkle Tree of DNN's Parameters*

## III. Experimentation

To test the viability of Merkle trees as a neural trojan detection technique, we created three steps in the experiment. The first step was to train a DNN classifier. We trained a trivial handwritten numeral classifier on a Jupyter Notebook platform. The classifier consisted of four layers: an input layer, hidden layers, and an output layer. Additionally, we implemented a SoftMax function. The DNN was trained and tested using the MNIST dataset. This dataset consisted of 70,000 images and labels of handwritten numerals ranging from 0 - 9. This dataset was divided into a 60,000-training dataset and a 10,000-testing dataset. Upon completion of training the model was saved to its checkpoint.

The second step was inserting the neural trojan. A malicious actor seeking to insert a neural trojan would require white-box access. To simulate this white-box access, it is assumed the malicious actor has gained access and elevated privileges through our stolen credentials. With their elevated privileges, the malicious actor is able to insert the single-bias trojan and save the trojaned model to the checkpoint. To test the validity of our single-bias attack, we attempted ten different single-bias attacks, each targeting a different sink class of the DNN.

The last step was the creation of a custom checkpoint. Upon loading the model from its checkpoint, two Merkle trees were created. The first was a Merkle tree of the saved model. The second was a Merkle tree of the model currently being used. If the root hashes did not match, the user would be alerted to the presence of the single-bias trojan, and the location of the neuron with the single bias would be reported by traversing both Merkle trees to discover the point of departure in node values.

## IV. Results and Discussion

Because the performance of the single-bias attack is based upon the probability of misclassifying a selected sink class and in the overall change in model accuracy, we recorded both the sparse categorical accuracy of each model and the probability of misclassifying the selected sink class. The baseline model had a sparse categorical accuracy of 0.9928, and the biased models' accuracies ranged from 0.8804 to 0.9025.

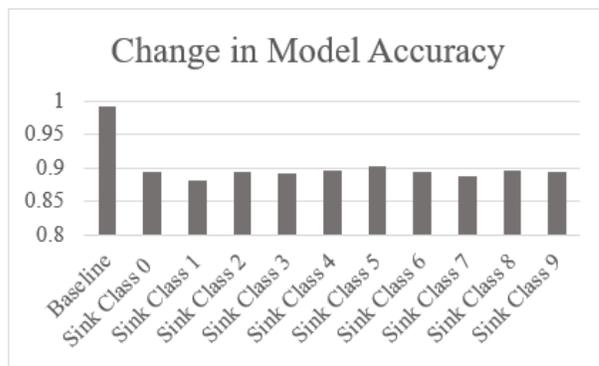

*Figure 1.3 Change in Model Accuracy*

At first glance, the reported change in accuracy appears unacceptable. Models with a ~10% degradation in accuracy are likely to be closely examined, resulting in the potential discovery of the single-bias attack. This large degradation in model accuracy is due to the small number of sink classes. The MNIST dataset is made up of ten different classes, and each class makes up between 9.04% and 11.24% of the data. If the trojaned model misclassifies the selected sink class with 100% accuracy, there will be a significant degradation of model performance because of the large proportion of a targeted sink class to the possible classes. For a DNN with a larger number of sink classes, such as facial recognition DNN, a trojaned DNN will experience a much smaller degradation in model accuracy, because there are many more possible sink classes to choose from.

The second criterion for the size of $\varepsilon$ in (1.2) was minimizing the probability of classifying input as the targeted sink class. By adding a significantly large $\varepsilon$, we were able to minimize the probability by 100%. This level of certainty is achievable because of the properties of the SoftMax function as described in (1.1). By ensuring that the neuron corresponding to the targeted sink class was the smallest in the output layer, the SoftMax function would never correctly label the targeted sink class.

To determine the effectiveness of the Merkle tree defense, we selected two criteria. First, for each of the 10 singl- bias attacks on the model, the Merkle tree defense must alert the user to the presence of a neural trojan. Second, for each of the 10 single bias attacks on the model, the Merkle tree defense must accurately report the location of the single-bias attack. For each of the 10 single bias attacks, the custom checkpoint with the Merkle tree accurately reported both the presence and location of the infected neuron.

## V. Conclusion

As DNNs are used in an increasing number of applications, a comprehensive look at defending DNNs is warranted. Comprehensive security of DNNs looks beyond solely defending against adversarial examples and data poisoning [8],[9]. It also includes defending against neural trojans. The current literature on defending DNNs from neural trojans is novel and interesting, but the proposed solutions are computationally expensive, not scalable, or commercially applicable. This research builds upon the capability of the hash-based alert system developed at the Naval Postgraduate School [5]. The Merkle tree defense can both alert the user to the presence of neural trojans and efficiently identify the infected neuron for future remediation. We hope this work will help lay the foundation for future researchers to develop comprehensive defenses of DNNs.

## **Author Information**

Joshua Strubel is a Sr. Computer Security Engineer at Savannah River National Laboratory, Ph.D. student at Liberty University, and adjunct instructor at Southern Wesleyan University.